\documentclass[twocolumn,showpacs,preprintnumbers,amsmath,amssymb]{revtex4}
\usepackage{graphicx}
\usepackage{amsmath}
\usepackage{appendix}
\usepackage{dcolumn}
\usepackage{makeidx}

\def\prl#1#2#3{{ Phys.   Rev.   Lett.  } {\bf #1}, #2 (#3)}

\def\pra#1#2#3{Phys.   Rev.   A {\bf #1}, #2 (#3)}

\def\rmp#1#2#3{Rev.   Mod.   Phys.   {\bf #1}, #2 (#3)}

\def\jsm#1#2#3#{J.Stat.Mech. {\bf#1}, #2 (#3)}

\def\noi{\noindent}
\def\bc{\begin{center}}
\def\ec{\end{center}}
 \newcommand{\bea}{\begin{equation}}
 \newcommand{\eea}{\end{equation}\noi}
 \newcommand{\ber}{\begin{eqnarray}}
 \newcommand{\eer}{\end{eqnarray}}

\begin{document}

\title{Wave-packet dynamics for a coupled Gross-Pitaevskii equation}
\author{Sukla Pal}\email{sukla@bose.res.in} 
\affiliation{Department of Theoretical Physics,\\S.N.Bose National Centre For Basic Sciences, JD-Block,Sector-III, Salt Lake City, Kolkata-700098, India}
\author{J. K. Bhattacharjee}\email{jkb@hri.res.in}
\affiliation{Harish-Chandra Research Institute, Chhatnag road, Jhunsi, Allahabad-211019, India}

\date{\today}

\begin{abstract}
Starting from coupled Gross-Pitaevskii equation (GPE) describing the dynamics of binary Bose Einstein Condensate (BEC), we have explored the dynamics of the width of initial wave-packets by following two different approaches. We have followed a recent coherent state based approach and have shown that in binary BEC the oscillation of the width of the wave packet can give rise to instability at a certain condition. Alternately we have proceeded by considering the trial ground state solutions for both the species and through the Ehrenfest theorem shown the existence of an instability which may lead to phase separation of the two species. 
\end{abstract}
 
\maketitle

\section{Introduction:}
The successful experimental achievements of BEC in 1995 in series of experiments of dilute atomic gases of Rb \cite {Ensher,pa}, Na \cite{Xerox}, Li \cite{N1} have created an interesting field of physics of cold dilute atomic gases. Both in theory and experiment this field of trapped BEC has revealed various interesting phenomena. Gross-Pitaevskii equation (GPE) was first developed to describe identical bosons by Gross \cite{a} and Pitaevskii \cite{b} in 1961, independently and since then it remains a widely used model in different areas of physics, such as quantum mechanics, condensed matter physics, nonlinear optics etc. After the enormous success of GPE in explaining almost all of the features of Bose Einstein condensate in the mean field level, coupled GPE has gained sufficient attraction to explain the binary species problem.

 The first two component condensate was produced with different hyperfine states of $Rb^{87}$. The experiments on $Rb^{87}$  atoms occupying the hyperfine states $|F=1,m_F=-1\rangle$ and $|F=2,m_f=1\rangle$ have been done by JILA group(Myatt et al.,1997). Later the group in MIT has been able to trap atoms in F=1 with different $m_F$ values using optical methods. Not only for same species with two different hyperfine states but also huge efforts have been given to prepare binary BEC with two different atomic species K-Rb \cite{k} and Cs-Li \cite{li}. Even the multicomponent BEC has been generating sufficient interest. Among the features of BEC in trapped gases are the spatial inhomogeneity created by the trapping potential \cite{6}, the interplay of coherent and nonlinear phenomena \cite{8,9,10}, collective oscillations \cite{11,12}, vortex formation\cite{13}, Josephson tunneling \cite{14,15} etc. Many such problems have already been addressed in recent literature by using the semi-classical theory introduced by Gross and Pitaevskii. In particular, the multispecies condensation and phase separation \cite{tune,16,3} have led to several theoretical and experimental investigations. In this work we will revisit the dynamics of the 1D binary BEC system, entirely from analytical perspective, using some interesting new insights that have been produced for the single component system \cite{1}.

Gaussian wave packets are minimum energy wave packets and are the coherent states for the simple harmonic oscillator in quantum mechanics. Recently these coherent states have been quite relevant in case of BEC \cite{1} as the existence of coherent state signifies the preservation of condensate without any diffusion of atoms from it. These coherent states are the states observed from the reference frame which are associated with the trajectory of classical and non-interacting particles. It has been found that the spreading of initial wave packet due to the repulsive nonlinear terms in GPE is strongly diminished by the harmonic trapping potential and shape of the wave packet remain unchanged with time \cite{1}. Quantum dynamics in presence of repulsive non-linearity in harmonic confining potential has become quite understandable by considering the coherent state initially. The existence of this coherent state in harmonic trapping potential for single species BEC serves as the motivation to investigate whether the formation of coherent state is possible or not in case of the multispecies problem. If it exists then an interesting offshoot can be the phase separation possibility in terms of coherent states. In this paper we have defined the coherent state for a coupled GPE and have shown the occurrence of an instability. Our article is organized as follows. The coherent state construction and the discussions regarding the stability of this state is carried out in sec II. In Sec III, we use the Ehrenfest point of view to arrive at qualitatively similar results. We conclude with a brief summary in sec IV.

\section{Coherent State Based Approach:}

In this section by following a recent coherent state based approach \cite{1} we will show how the concept of coherent state can be generalized in case of two species BEC. We consider the coupled Gross Pitaevskii equation with identical masses $(m)$ and chemical potentials ($\mu$) of the species $\alpha$ and $\beta$ but different repulsive non-linear couplings ($g_{\alpha}$, $g_{\beta}$ and $g_{\alpha\beta}$).

\begin{eqnarray} \label{eq1}
\begin{aligned}
i\hbar\partial_t\psi_{\alpha}(x,t)=-\frac{\hbar^2}{2m}\nabla^2\psi_{\alpha}(x,t)+V_{ext}(x)\psi_{\alpha}(x,t)\\+\Big(g_{\alpha}|\psi_{\alpha}|^2+g_{\alpha\beta}|\psi_{\beta}|^2-\mu\Big )\psi_{\alpha}\\
i\hbar\partial_t\psi_{\beta}(x,t)=-\frac{\hbar^2}{2m}\nabla^2\psi_{\beta}(x,t)+V_{ext}(x)\psi_{\beta}(x,t)\\+\Big(g_{\beta}|\psi_{\beta}|^2+g_{\alpha\beta}|\psi_{\alpha}|^2-\mu\Big )\psi_{\beta},
\end{aligned}
\end{eqnarray} 
The two species are confined in the harmonic potential ($V_{ext}(x)=\frac{1}{2}m\omega^2)$ with normalization $\int|\psi_{\alpha}(x)|^2dx=N_{\alpha}$, $\int|\psi_{\beta}(x)|^2dx=N_{\beta}$, where $N_{\alpha}$ and $N_{\beta}$ are the total number of species $\alpha$ and $\beta$ respectively. The coupling constants are $g_{\alpha}=\frac{4\pi\hbar^2a_{\alpha}}{m}$, $g_{\beta}=\frac{4\pi\hbar^2a_{\beta}}{m}$, $g_{\alpha\beta}=\frac{4\pi\hbar^2a_{\alpha\beta}}{m}$, where $a_{\alpha}$ and $a_{\beta}$ are the s-wave scattering length of the species $\alpha$ and $\beta$ respectively, while $a_{\alpha\beta}$ is determined when an atom of species $\alpha$ is scattered by one of species $\beta$. Since we are searching for a coherent state, we need to find a form of wave packet which remains of the same shape in course of evolution. Hence the following can be a trial solution of Eq. (\ref{eq1}):
\ber\label{eq2}
\begin{aligned}
\psi_{\alpha}(x,t)=\phi _{\alpha}(x-x_{0\alpha}(t),t)e^{\frac{ip_{0\alpha}(t)}{\hbar}\Big(x-\frac{x_{0\alpha}(t)}{2}\Big)}\\
\psi_{\beta}(x,t)=\phi _{\beta}(x-x_{0\beta}(t,t))e^{\frac{ip_{0\beta}(t)}{\hbar}\Big(x-\frac{x_{0\beta}(t)}{2}\Big)}
\end{aligned}
\eer
  
These solutions represent traveling wave packets being translated along the phase space which is 4 dimensional in our case. The phase space trajectory of $(x_{0\alpha}(t),x_{0\beta}(t),p_{0\alpha}(t),p_{0\beta}(t))$ will act as the reference frame for the coherent states if it coincides with the phase-space trajectory of a classical, non-interacting particle. This constraint dictates that the trajectory will be governed by the following sets of dynamical equations:
\bea\label{eq3}
\begin{aligned}
{\dot{x}}_{0\alpha}(t) &= \frac{p_{0\alpha}(t)}{m}\\
{\dot{x}}_{0\beta}(t) &= \frac{p_{0\beta}(t)}{m}\\
{\dot{p}}_{0\alpha}(t) &= -\frac{\partial V_{ext}}{\partial x}\Big|_{x_{0\alpha}(t)}\\
{\dot{p}}_{0\beta}(t) &= -\frac{\partial V_{ext}}{\partial x}\Big|_{x_{0\beta}(t)}
\end{aligned}
\eea
  
Following the same arguments as in \cite{1} in case of binary species, we can conclude that from this special reference frame as described above, the quantum equation of motions for $\alpha$ and $\beta$ take the following form in a harmonic potential. 

\ber\label{eq4}
\begin{aligned}
i\hbar\partial_t\phi_{\alpha}=-\frac{\hbar^2}{2m}\nabla^2\phi_{\alpha}+V_{ext}(x)\phi_{\alpha}\\+\Big(g_{\alpha}|\phi_{\alpha}|^2+g_{\alpha\beta}|\phi_{\beta}|^2-\mu\Big )\phi_{\alpha}\\
i\hbar\partial_t\phi_{\beta}=-\frac{\hbar^2}{2m}\nabla^2\phi_{\beta}+V_{ext}(x)\phi_{\beta}\\+\Big(g_{\beta}|\phi_{\beta}|^2+g_{\alpha\beta}|\phi_{\alpha}|^2-\mu\Big )\phi_{\beta},
\end{aligned}
\eer
In \cite{1} the argument is given for single species but this argument can remain valid for any multispecies problem as long as one is dealing with the trajectory of non-interacting particles. The coherent states of coupled GPE can be described by the stationary states of coupled GPE governed by the equations below.
\begin{equation} \label{eq5}
\begin{aligned}
\Big(-\frac{\hbar^2}{2m}\nabla^2+V_{ext}+g_{\alpha}|\phi_{0\alpha}|^2+g_{\alpha\beta}|\phi_{0\beta}|^2\Big )\phi_{0\alpha} &=\mu\phi_{0\alpha}\\
\Big(-\frac{\hbar^2}{2m}\nabla^2+V_{ext}+g_{\beta}|\phi_{0\beta}|^2+g_{\alpha\beta}|\phi_{0\alpha}|^2\Big )\phi_{0\beta} &=\mu\phi_{0\beta},
\end{aligned}
\end{equation} 
Where,
\bea \label{eq6}
\begin{aligned}
\psi_{0\alpha}(x,t)=\phi_{0\alpha}(x-x_{0\alpha}(t))e^{\frac{ip_{0\alpha}(t)}{\hbar}\Big(x-\frac{x_{0\alpha}(t)}{2}\Big)}\\
\psi_{0\beta}(x,t)=\phi_{0\beta}(x-x_{0\beta}(t))e^{\frac{ip_{0\beta}(t)}{\hbar}\Big(x-\frac{x_{0\beta}(t)}{2}\Big)}
\end{aligned}
\eea
are the time dependent exact solution of coupled GPE given in Eq. (\ref{eq1}).\\

We now need to study the stability of the coherent state against shape deformation. We perturb the coherent state solution $\phi_{0\alpha}$ and $\phi_{0\beta}$ and are interested to know the stability of the time dependent solution given by Eq. (\ref{eq6}). For this we take, $\phi_{\alpha}=\phi_{0\alpha}+\delta\phi_{\alpha}$ and $\phi_{\beta}=\phi_{0\beta}+\delta\phi_{\beta}$. Substituting this expressions in Eq. (\ref{eq4}) we get the following pair of equations:
\ber
i\hbar\partial_t\delta\phi_{\alpha}&=&\Big(-\frac{\hbar^2}{2m}\nabla^2+\tilde{V}_{ext}+2g_{\alpha}|\phi_{0\alpha}|^2+g_{\alpha\beta}|\phi_{0\beta}|^2\Big )\delta\phi_{\alpha}\nonumber\\&+&g_{\alpha}\phi_{0\alpha}^2\delta\phi_{\alpha}^{\ast}+g_{\alpha\beta}\phi_{0\alpha}\big(\phi_{0\beta}^{\ast}\delta\phi_{\beta}+\phi_{0\beta}\delta\phi_{\beta}^{\ast}\big)\label{eq7}\\
i\hbar\partial_t\delta\phi_{\beta}&=&\Big(-\frac{\hbar^2}{2m}\nabla^2+\tilde{V}_{ext}+2g_{\beta}|\phi_{0\beta}|^2+g_{\alpha\beta}|\phi_{0\alpha}|^2\Big )\delta\phi_{\beta}\nonumber\\&+&g_{\beta}\phi_{0\beta}^2\delta\phi_{\beta}^{\ast}+g_{\alpha\beta}\phi_{0\beta}\big(\phi_{0\alpha}^{\ast}\delta\phi_{\alpha}+\phi_{0\alpha}\delta\phi_{\alpha}^{\ast}\big)\label{eq8},
\eer
where, $\tilde{V}_{ext}=V_{ext}-\mu$ and we are taking real $\phi_{0\alpha}$ and $\phi_{0\beta}$ such that $\phi_{0\alpha}^2=|\phi_{0\alpha}|^2=n_{\alpha}$ and  $\phi_{0\beta}^2=|\phi_{0\beta}|^2=n_{\beta}$ respectively. Considering $\delta\phi_{\alpha}=\sum_i\Big(u_{i\alpha}(x)e^{-i\omega_it}-v_{i\alpha}^{\ast}(x)e^{i\omega_it}\Big)$ and $\delta\phi_{\beta}=\sum_i\Big(u_{i\beta}(x)e^{-i\omega_it}-v_{i\beta}^{\ast}(x)e^{i\omega_it}\Big)$ and keeping only the linear term in $\delta\phi_{\alpha}$ and $\delta\phi_{\beta}$, we obtain the following sets of equations for species $\alpha$ from (\ref{eq7}) and its complex conjugate.

\begin{eqnarray}
(1+2\xi\epsilon_i)u_{i\alpha}=-\xi^2\Delta u_{i\alpha}+x^2u_{i\alpha}+\bar{n_{\alpha}}(2u_{i\alpha}-v_{i\alpha})\nonumber\\+c_{\beta}\bar{n_{\beta}}u_{i\alpha}+c_{\beta}\sqrt{\frac{g_{\beta}}{g_{\alpha}}}\bar{n}_{\beta}\sqrt{\frac{\bar{n_\alpha}}{\bar{n_\beta}}}(u_{i\beta}-v_{i\alpha})\label{eq9}\\
(1-2\xi\epsilon_i)v_{i\alpha}=-\xi^2\Delta v_{i\alpha}+x^2v_{i\alpha}+\bar{n_{\alpha}}(2v_{i\alpha}-u_{i\alpha})\nonumber\\+c_{\beta}\bar{n_{\beta}}v_{i\alpha}+c_{\beta}\sqrt{\frac{g_{\beta}}{g_{\alpha}}}\bar{n}_{\beta}\sqrt{\frac{\bar{n_\alpha}}{\bar{n_\beta}}}(v_{i\beta}-u_{i\alpha})\label{eq10}
\end{eqnarray}

To obtain Eq. (\ref{eq9}) we re-scaled $x$ as $\tilde{x}=\frac{x}{l}$ where $l=\big(\frac{2\mu}{m\omega^2}\big)^{1/2}$ is characteristic size of the condensate. We define a parameter $\xi=\frac{\hbar\omega}{2\mu}$ and under Thomas Fermi approximation  $\xi\ll1$ \cite{2}. We write $\hbar\omega_i=\epsilon_i\hbar\omega$ and we have defined $\frac{g_{\alpha}n_{\alpha}}{\mu}=\bar{n}_{\alpha}$, $\frac{g_{\beta}n_{\beta}}{\mu}=\bar{n}_{\beta}$, $\frac{g_{\alpha\beta}n_{\alpha}}{\mu}=C_{\alpha}\bar{n}_{\alpha}$ and $\frac{g_{\alpha\beta}n_{\beta}}{\mu}=C_{\beta}\bar{n}_{\beta}$ where, $C_{\beta}=\frac{g_{\alpha\beta}}{g_{\beta}}$ and $C_{\alpha}=\frac{g_{\alpha\beta}}{g_{\alpha}}$. For simplicity of notation after re-scaling in the above way we put $x$ in place of $\tilde{x}$. We write $\frac{\partial^2}{\partial x^2}=\Delta$ and $\hbar\omega_i=\epsilon_i\hbar\omega$.\\

Adding and subtracting Eq. (\ref{eq9}) and Eq. (\ref{eq10}) and considering $u_{i\alpha}\pm v_{i\alpha}=f_{i\alpha}^{pm}$, we get the following equations:
\ber
f_{i\alpha}^{+}+2\xi\epsilon_if_{i\alpha}^{-}=\big(-\xi^2\Delta+x^2+\bar{n}_{\alpha}+C_{\beta}\bar{n}_{\beta}\big)f_{i\alpha}^{+}\label{eq11}\\
f_{i\alpha}^{-}+2\xi\epsilon_if_{i\alpha}^{+}=\big(-\xi^2\Delta+x^2+3\bar{n}_{\alpha}+C_{\beta}\bar{n}_{\beta}\big)f_{i\alpha}^{-}\nonumber\\+2C_{\beta}\sqrt{\frac{g_{\beta}}{g_{\alpha}}}\bar{n}_{\beta}\sqrt{\frac{\bar{n}_{\alpha}}{\bar{n}_{\beta}}}f_{i\alpha}^{-}\label{eq12}
\eer
After doing the similar calculation for species $\beta$, we find following equations:
\ber
f_{i\beta}^{+}+2\xi\epsilon_if_{i\beta}^{-}=\big(-\xi^2\Delta+x^2+\bar{n}_{\beta}+C_{\alpha}\bar{n}_{\alpha}\big)f_{i\beta}^{+}\label{eq101}\label{eq13}\\
f_{i\beta}^{-}+2\xi\epsilon_if_{i\beta}^{+}=\big(-\xi^2\Delta+x^2+3\bar{n}_{\alpha}+C_{\alpha}\bar{n}_{\alpha}\big)f_{i\beta}^{-}\nonumber\\+2C_{\alpha}\sqrt{\frac{g_{\alpha}}{g_{\beta}}}\bar{n}_{\alpha}\sqrt{\frac{\bar{n}_{\beta}}{\bar{n}_{\alpha}}}f_{i\beta}^{-}\label{eq102}\label{eq14}
\eer

After doing few steps and neglecting the terms containing $\xi^2$ and considering $f_{i\alpha}^{\pm}=(1-x^2)^{\pm1/2}\eta_{\alpha}(x)$ and $f_{i\beta}^{\pm}=(1-x^2)^{\pm1/2}\eta_{\beta}(x)$ we get finally the following two coupled equations:
\bea\label{eq15}
\begin{aligned}
(1-x^2)\frac{d^2\eta_{\alpha}}{dx^2}-2x\frac{d\eta_{\alpha}}{dx}+C_{\beta}\sqrt{\frac{1-C_{\alpha}}{1-C_{\beta}}}\sqrt{\frac{g_{\beta}}{g_{\alpha}}}\big[(1-x^2)\frac{d^2\eta_{\beta}}{dx^2}\\-2x\frac{d\eta_{\beta}}{dx}\big]+\frac{2}{1-C_{\beta}}\epsilon_i^2\eta_{\alpha}=0\\
(1-x^2)\frac{d^2\eta_{\beta}}{dx^2}-2x\frac{d\eta_{\beta}}{dx}+C_{\alpha}\sqrt{\frac{1-C_{\beta}}{1-C_{\alpha}}}\sqrt{\frac{g_{\alpha}}{g_{\beta}}}\big[(1-x^2)\frac{d^2\eta_{\alpha}}{dx^2}\\-2x\frac{d\eta_{\alpha}}{dx}\big]+\frac{2}{1-C_{\alpha}}\epsilon_i^2\eta_{\beta}=0\\
\end{aligned}
\eea
 If we consider $g_{\alpha\beta}=0$ in any of the equations given in Eq. (\ref{eq15}), we recover the equation for single species BEC \cite{2}:
\ber
(1-x^2)\frac{d^2\eta}{dx^2}-2x\frac{d\eta}{dx}+2\epsilon_i^2\eta=0\label{eq16}
\eer
Eq. (\ref{eq16}) leads to the well known dispersion relation for single species
\bea\label{eq17}
E_i=\hbar\omega_i=\hbar\omega\sqrt{\frac{n(n+1)}{2}}
\eea
 where $n=1$ for ground state solution.\\

Consider the special case $g_{\alpha}=g_{\beta}$, so that $C_{\alpha}=C_{\beta}=C$, in our binary species problem, we find by adding the pair of equations in Eq. (\ref{eq15}):
\ber\label{eq18}
(1-x^2)\frac{d^2\theta}{dx^2}-2x\frac{d\theta}{dx}+\frac{2\epsilon^2}{1-C^2}\theta=0
\eer
where, $\theta=\eta_{\alpha}+\eta_{\beta}$. The eigen values are now given by $\omega_i=\omega\sqrt{1-C^2}\sqrt{\frac{n(n+1)}{2}}$. We note that $\omega_i$ vanishes at $C=1$ signaling an instability. This implies $g_{\alpha\beta}=g_{\alpha}$. It is important to note that our instability criterion is consistent with
\ber\label{eq19}
 g_{\alpha\beta}>\sqrt{g_{\alpha}g_{\beta}}
\eer
Condition (\ref{eq19}) is well known in literature for causing the phase separation in binary BEC \cite{tune} and here we have derived the condition from coherent state based approach. 


\section{Evolution of the wave packet through Ehrenfest like approach:}

We begin with the 1D coupled time dependent GPE of the following form:
\begin{eqnarray}\label{eq20}
\begin{aligned}
i\hbar\partial_t\psi_{\alpha}=\Big(-\frac{\hbar^2}{2m}\nabla^2+V_{ext}+g_{\alpha}|\psi_{\alpha}|^2+g_{\alpha\beta}|\psi_{\beta}|^2\Big )\psi_{\alpha}\\
i\hbar\partial_t\psi_{\beta}=\Big(-\frac{\hbar^2}{2m} \nabla^2+V_{ext}+g_{\beta}|\psi_{\beta}|^2+g_{\alpha\beta}|\psi_\alpha|^2\Big )\psi_{\beta} 
\end{aligned}
\end{eqnarray} 
The normalization, external confining potential and the coupling constants have the same form as taken in the previous section. For $V_{ext}(x)=\frac{1}{2}\omega^2x^2$, appropriate rescalings can be done to write Eqs. (\ref{eq20}) in the dimensionless form:
\ber\label{eq21}
\begin{aligned}
i\partial_t\psi_{\alpha}&=&\Big(-\frac{1}{2}\nabla^2+\frac{1}{2}\Omega^2x^2+g_{\alpha}|\psi_{\alpha}|^2+g_{\alpha\beta}|\psi_{\beta}|^2\Big )\psi_{\alpha}\\
i\partial_t\psi_{\beta}&=&\Big(-\frac{1}{2}\nabla^2+\frac{1}{2}\Omega^2x^2+g_{\beta}|\psi_{\beta}|^2+g_{\alpha\beta}|\psi_{\alpha}|^2\Big )\psi_{\beta}
\end{aligned}
\eer
 We now consider the following trial ground state solution for Eq. (\ref{eq21}). 
\ber\label{eq22}
\begin{aligned}
\psi_{\alpha}(x,t)=\frac{\sqrt{N_{\alpha}}}{(2\pi W_{\alpha}^2)^{1/4}}e^{-\frac{(x-x_{0\alpha}(t))^2}{4W_{\alpha}^2}}e^{ip_{0\alpha}(x-\frac{x_{0\alpha}}{2})}\\
\psi_{\beta}(x,t)=\frac{\sqrt{N_{\beta}}}{(2\pi W_{\beta}^2)^{1/4}}e^{-\frac{(x-x_{0\beta}(t))^2}{4W_{\beta}^2}}e^{ip_{0\beta}(x-\frac{x_{0\beta}}{2})}
\end{aligned}
\eer
The expectation value is defined as usual as $\langle x\rangle_{\alpha}=\int x\psi_{\alpha}(x)\psi_{\alpha}^{\ast}dx$. Differentiating with respect to time:
\ber\label{eq23}
\begin{aligned}
\dot{x}_{0\alpha}=\langle \dot{x}\rangle_{\alpha}=\int[\frac{\partial\psi_{\alpha}}{\partial t}x\psi^{\ast}_{\alpha}+\frac{\partial\psi^{\ast}_{\alpha}}{\partial t}x\psi_{\alpha}]dx\\
\dot{x}_{0\beta}=\langle \dot{x}\rangle_{\beta}=\int[\frac{\partial\psi_{\beta}}{\partial t}x\psi^{\ast}_{\beta}+\frac{\partial\psi^{\ast}_{\beta}}{\partial t}x\psi_{\beta}]dx
\end{aligned}
\eer
With the help of Eq. (\ref{eq21}), the above reduce to
\ber\label{eq24}
\begin{aligned}
N_{\alpha}\dot{x}_{0\alpha}&=&\frac{i}{2}\int\Big(\psi_{\alpha}\partial_x\psi_{\alpha}^{\ast}-\psi_{\alpha}^{\ast}\partial_x\psi_{\alpha}\Big)dx\\
N_{\beta}\dot{x}_{0\beta}&=&\frac{i}{2}\int\Big(\psi_{\beta}\partial_x\psi_{\beta}^{\ast}-\psi_{\beta}^{\ast}\partial_x\psi_{\beta}\Big)dx
\end{aligned}
\eer
Applying Eq. (\ref{eq22}) in Eq. (\ref{eq24}) we obtain the Ehrenfest equations:
\ber\label{eq25}
\begin{aligned}
\dot{x}_{0\alpha}= p_{0\alpha}\\
\dot{x}_{0\beta}= p_{0\beta}
\end{aligned}
\eer
Similarly for $\langle p\rangle$, we have $\frac{d}{dt}\langle p\rangle=\frac{d}{dt}\langle \psi^{\ast}(x)\big(-i\frac{d}{dx}\big)\psi(x)\rangle$. With the help of Eq. (\ref{eq21}) and Eq. (\ref{eq22}) we arrive at the following pair of equations.
\ber\label{eq26}
\begin{aligned}
\dot{p_{0\alpha}}=-\Omega^2x_{0\alpha}+\frac{N_\beta}{\sqrt{2\pi}}\frac{g_{\alpha\beta}\Delta x}{W^3}e^{-\frac{\Delta x^2}{2W^2}}\\
\dot{p_{0\beta}}=-\Omega^2x_{0\beta}-\frac{N_\alpha}{\sqrt{2\pi}}\frac{g_{\alpha\beta}\Delta x}{W^3}e^{-\frac{\Delta x^2}{2W^2}}
\end{aligned}
\eer
where, $x_{0\alpha}-x_{0\beta}=\Delta x$ and $(W_{\alpha}^2+W_{\beta}^2)^{1/2}=W$.
We also have $\langle x^2\rangle=\int x^2\psi^{\ast}\psi dx$ and taking time derivatives twice (details are described in Appendix A):
\ber\label{eq29}
\begin{aligned}
W_{\alpha}\ddot{W_{\alpha}}+\dot{W_{\alpha}^2}=\frac{1}{4W_{\alpha}^2}-\Omega^2W_{\alpha}^2+\frac{g_{\alpha}}{4\sqrt{\pi}}\frac{N_\alpha}{W_{\alpha}}\\+\frac{g_{\alpha\beta}}{2}\frac{N_{\beta}}{\sqrt{2\pi W^2}}e^{-\frac{\Delta x^2}{2W^2}}\Big[1+2\frac{x_{0\alpha}\Delta x}{W^2}\Big]\\
W_{\beta}\ddot{W_{\beta}}+\dot{W_{\beta}^2}=\frac{1}{4W_{\beta}^2}-\Omega^2W_{\beta}^2+\frac{g_{\beta}}{4\sqrt{\pi}}\frac{N_\beta}{W_{\beta}}\\+\frac{g_{\alpha\beta}}{2}\frac{N_{\alpha}}{\sqrt{2\pi W^2}}e^{-\frac{\Delta x^2}{2W^2}}\Big[1+2\frac{x_{0\beta}\Delta x}{W^2}\Big]
\end{aligned}
\eer
In absence of $g$'s, the equilibrium width for both the species are equal and given by: $W_{\alpha}|_{eq}=(\frac{1}{4\Omega^2})^{1/2}=W_{\beta}|_{eq}=W_{eq}$. Defining, $\sigma_{\alpha}=\frac{W_{\alpha}}{W_{eq}}$, $\sigma_{\beta}=\frac{W_{\beta}}{W_{eq}}$, $\sigma=\frac{W}{W_{eq}}=(\sigma_{\alpha}^2+\sigma_{\beta}^2)^{1/2}$. Rewriting Eq. (\ref{eq29}) in terms of $\sigma$'s:
\ber\label{eq30}
\begin{aligned}
\sigma_{\alpha}\ddot{\sigma_{\alpha}}+\dot{\sigma_{\alpha}^2} = 2\Omega^2\Big[\frac{1}{\sigma_{\alpha}^2}-\sigma_{\alpha}^2+\frac{g_{\alpha N_\alpha}}{\sqrt{2\pi\Omega\sigma_{\alpha}^2}}\\+\frac{g_{\alpha\beta}N_{\beta}}{\sqrt{\pi\Omega\sigma^2}}e^{-\frac{\Delta x^2}{2W^2}}(1+4\Omega\frac{x_{0\alpha}\Delta x}{\sigma^2})\Big]\\
\sigma_{\beta}\ddot{\sigma_{\beta}}+\dot{\sigma_{\beta}^2} =2\Omega^2\Big[\frac{1}{\sigma_{\beta}^2}-\sigma_{\beta}^2+\frac{g_{\beta N_\beta}}{\sqrt{2\pi\Omega\sigma_{\beta}^2}}\\+\frac{g_{\alpha\beta}N_{\alpha}}{\sqrt{\pi\Omega\sigma^2}}e^{-\frac{\Delta x^2}{2W^2}}(1+4\Omega\frac{x_{0\beta}\Delta x}{\sigma^2})\Big]
\end{aligned}
\eer
We can consider three interesting cases:
\subsection{$x_{0\alpha}=x_{0\beta}$ :}

Under this condition Eq. (\ref{eq30}) reduces to the following form:
\ber\label{eq31}
\begin{aligned}
\sigma_{\alpha}\ddot{\sigma_{\alpha}}+\dot{\sigma_{\alpha}^2} &=& 2\Omega^2\Big[\frac{1}{\sigma_{\alpha}^2}-\sigma_{\alpha}^2+\frac{g_{\alpha}N_{\alpha}}{\sqrt{2\pi\Omega\sigma_{\alpha}^2}}+\frac{g_{\alpha\beta}N_{\beta}}{\sqrt{\pi\Omega\sigma^2}}\Big]\\
\sigma_{\beta}\ddot{\sigma_{\beta}}+\dot{\sigma_{\beta}^2} &=& 2\Omega^2\Big[\frac{1}{\sigma_{\beta}^2}-\sigma_{\beta}^2+\frac{g_{\beta N_\beta}}{\sqrt{2\pi\Omega\sigma_{\beta}^2}}+\frac{g_{\alpha\beta}N_{\alpha}}{\sqrt{\pi\Omega\sigma^2}}\Big]
\end{aligned}
\eer
We will do a linear stability analysis for both the species around the equilibrium width $\sigma_{\alpha}|_{eq}$ and $\sigma_{\beta}|_{eq}$:
\begin{eqnarray*}
\sigma_{\alpha}(t)=\sigma_{\alpha}|_{eq}(g)+\delta\sigma_{\alpha}(t)\\
\sigma_{\beta}(t)=\sigma_{\beta}|_{eq}(g)+\delta\sigma_{\beta}(t)
\end{eqnarray*}
Keeping only the linear terms in $\delta\sigma_{\alpha}$ and $\delta\sigma_{\alpha}$, we obtain the following two coupled differential equation:

\ber\label{eq32}
\begin{aligned}
\ddot{\delta\sigma_{\alpha}}(t)+\omega_{\alpha 1}\delta\sigma_{\alpha}(t)+\omega_{\alpha 2}\delta\sigma_{\beta}(t)=0\\
\ddot{\delta\sigma_{\beta}}(t)+\omega_{\beta 1}\delta\sigma_{\beta}(t)+\omega_{\beta 2}\delta\sigma_{\alpha}(t)=0
\end{aligned}
\eer
Where, 
\ber\label{eq33}
\begin{aligned}
\omega_{\alpha 1}=\Omega^2[2+\frac{2}{\sigma_{\alpha}|_{eq}^4}+\frac{g_{\alpha}N_{\alpha}}{\sqrt{2\pi\Omega}}\frac{1}{\sigma_{\alpha}|_{eq}^3}\\+\frac{g_{\alpha\beta}N_{\beta}}{\sqrt{\pi\Omega}}\frac{1}{(\sigma_{\alpha}|_{eq}^2+\sigma_{\beta}|_{eq}^2)^{3/2}}]\\
\omega_{\alpha 2}=\Omega^2(\frac{\sigma_{\beta|_{eq}}}{\sigma_{\alpha}|_{eq}}\frac{g_{\alpha\beta}N_{\beta}}{\sqrt{\pi\Omega}}\frac{1}{(\sigma_{\alpha}|_{eq}^2+\sigma_{\beta}|_{eq}^2)^{3/2}})\\
\omega_{\beta 1}=\Omega^2[2+\frac{2}{\sigma_{\beta}|_{eq}^4}+\frac{g_{\beta}N_{\beta}}{\sqrt{2\pi\Omega}}\frac{1}{\sigma_{\beta}|_{eq}^3}\\+\frac{g_{\alpha\beta}N_{\alpha}}{\sqrt{\pi\Omega}}\frac{1}{(\sigma_{\alpha}|_{eq}^2+\sigma_{\beta}|_{eq}^2)^{3/2}}]\\
\omega_{\beta 2}=\Omega^2(\frac{\sigma_{\alpha|_{eq}}}{\sigma_{\beta}|_{eq}}\frac{g_{\alpha\beta}N_{\alpha}}{\sqrt{\pi\Omega}}\frac{1}{(\sigma_{\beta}|_{eq}^2+\sigma_{\alpha}|_{eq}^2)^{3/2}})
\end{aligned}
\eer
So the width of both the species will perform coupled oscillation around their equilibrium width. To check the stability of the coupled oscillation, we need to find the normal mode frequencies $\omega$ from:
\begin{displaymath}
\left|\begin{array}{cc}
-\omega^2+\omega_{\alpha 1} & \omega_{\alpha 2} \\
 \omega_{\beta 2} & -\omega^2+\omega_{\beta 1}\\
\end{array} \right|=0\nonumber\\
\end{displaymath}
The frequencies are:
\ber\label{eq34}
2\omega^2=\omega_{\alpha 1}+\omega_{\beta 1}\pm \sqrt{(\omega_{\alpha 1}+\omega_{\beta 1})^2-4(\omega_{\alpha 1}\omega_{\beta 1}-\omega_{\alpha 2}\omega_{\beta 2})}\nonumber\\
\eer

The above expression suggests that if $\omega_{\alpha 1}\omega_{\beta 1}<\omega_{\alpha 2}\omega_{\beta 2}$, $\Delta x=0$ condition can in principal gives rise to instability. In another words, though both the wave packets are at the same position, the widths need not be oscillatory and there can be a shape instability in the system. 
\subsection{$x_{0\alpha}\ne x_{0\beta}$ with constant $(p_{0\alpha}-p_{0\beta})$:}

This implies that at long time limit ($\Delta t\rightarrow \infty$), $e^{-\Delta x}=e^{-\Delta t(p_{\alpha}-p_{\beta})}\simeq 0$. Hence in this case Eq. (\ref{eq30}) reduces to:
\ber\label{eq35}
\begin{aligned}
\sigma_{\alpha}\ddot{\sigma_{\alpha}}+\dot{\sigma_{\alpha}^2} = 2\Omega^2\Big[\frac{1}{\sigma_{\alpha}^2}-\sigma_{\alpha}^2+\frac{g_{\alpha}N_{\alpha}}{\sqrt{2\pi\Omega\sigma_{\alpha}^2}}\Big]\\
\sigma_{\beta}\ddot{\sigma_{\beta}}+\dot{\sigma_{\beta}^2} = 2\Omega^2\Big[\frac{1}{\sigma_{\beta}^2}-\sigma_{\beta}^2+\frac{g_{\beta N_\beta}}{\sqrt{2\pi\Omega\sigma_{\beta}^2}}\Big]
\end{aligned}
\eer
So after a long time, the width equations for both the species are decoupled and the width of each of the condensate will fluctuate around their mean equilibrium value independently obeying the following equations .
\ber\label{eq36}
\begin{aligned}
\delta\ddot{\sigma_{\alpha}}+\tilde{\omega_{\alpha}}\delta\sigma_{\alpha}(t)=0\\
\delta\ddot{\sigma_{\beta}}+\tilde{\omega_{\beta}}\delta\sigma_{\beta}(t)=0
\end{aligned}
\eer
Where, 
\begin{eqnarray*}
\tilde{\omega_{\alpha }}=\Omega^2\big[2+\frac{2}{\sigma_{\alpha}|_{eq}^4}+\frac{g_{\alpha}N_{\alpha}}{\sqrt{2\pi\Omega}}\frac{1}{\sigma_{\alpha}|_{eq}^3}\big]\\
\tilde{\omega_{\beta }}=\Omega^2\big[2+\frac{2}{\sigma_{\beta}|_{eq}^4}+\frac{g_{\beta}N_{\beta}}{\sqrt{2\pi\Omega}}\frac{1}{\sigma_{\beta}|_{eq}^3}\big]
\end{eqnarray*}
\subsection{$x_{0\alpha}\ne x_{0\beta}$ with variable $p_{0\alpha}$ and $p_{0\beta}$:}

With the help of Eq. (\ref{eq25} and Eq. (\ref{eq26}) we get the following set of dynamical equations:
\ber\label{eq37}
\begin{aligned}
\ddot{x_{0\alpha}}=\dot{p_{0\alpha}}=-\Omega^2x_{0\alpha}+\frac{N_{\beta}}{\sqrt{2\pi}}\frac{g_{\alpha\beta}\Delta x}{W^3}e^{-\frac{\Delta x^2}{2W^2}}\\
\ddot{x_{0\beta}}=\dot{p_{0\beta}}=-\Omega^2x_{0\beta}-\frac{N_{\alpha}}{\sqrt{2\pi}}\frac{g_{\alpha\beta}\Delta x}{W^3}e^{-\frac{\Delta x^2}{2W^2}}
\end{aligned}
\eer
The fixed points of the above system are $x_{0\alpha}^{\ast}$ and $x_{0\beta}^{\ast}$ and are to be obtained by equating to zero the rhs. of Eq. (\ref{eq37}).

 We now consider the stability. By writing $x_{0\alpha}=x_{0\alpha}^{\ast}+\delta x_{0\alpha}$ and $x_{0\beta}=x_{0\beta}^{\ast}+\delta x_{0\beta}$ and doing a linear stability analysis we find the following structure.

\begin{displaymath}
\left(\begin{array}{c}\label{eq38}
\delta x_{0\alpha}\\
\delta x_{0\beta}\\
\end{array}\right)
\left(\begin{array}{cc}
-\Omega^2+N_{\beta}\tilde{g} & -N_{\beta}\tilde{g}\\
-N_{\alpha}\tilde{g} & -\Omega^2+N_{\alpha}\tilde{g}\\
\end{array} \right)
=\left(\begin{array}{c}
\delta\ddot{x}_{0\alpha}\\
\delta\ddot{x}_{0\beta}\\
\end{array}\right)
\end{displaymath}
where, $\tilde{g}=\frac{1}{\sqrt{2\pi}}\frac{g_{\alpha\beta}}{W^3}$. The eigen values of the coefficient matrix given above are $\lambda_{\pm}=\frac{1}{2}[(\tilde{g}N-2\Omega^2)\pm \tilde{g}N]$ where $N=N_{\alpha}+N_{\beta}$. Instability can occur in the system if
\ber
N &>& \frac{\Omega^2}{\tilde{g}}\nonumber\\
g_{\alpha\beta} &>& \sqrt{2\pi}\frac{\Omega^2}{N}(W_{\alpha}^2+W_{\beta}^2)\label{eq39}
\eer
To understand the nature of the instability we note that from Eq. (\ref{eq37}), on subtracting one from the other, 
\ber
\ddot{\Delta x} &=&-\frac{dV_{eff}}{dx}\nonumber
\eer
where
\ber\label{eq40}
V_eff=\frac{\Omega^2}{2}(\Delta x)^2+\frac{N}{\sqrt{2\pi}}\frac{g_{\alpha\beta}}{W}e^{-\frac{\Delta x^2}{2W^2}}
\eer
Now if we plot this effective potential with $g_{\alpha\beta}$ as parameter we find that the effective potential is simple harmonic when $g_{\alpha\beta}=0$.
\begin{figure}[!htbp]
\includegraphics[angle=0,scale=0.7]{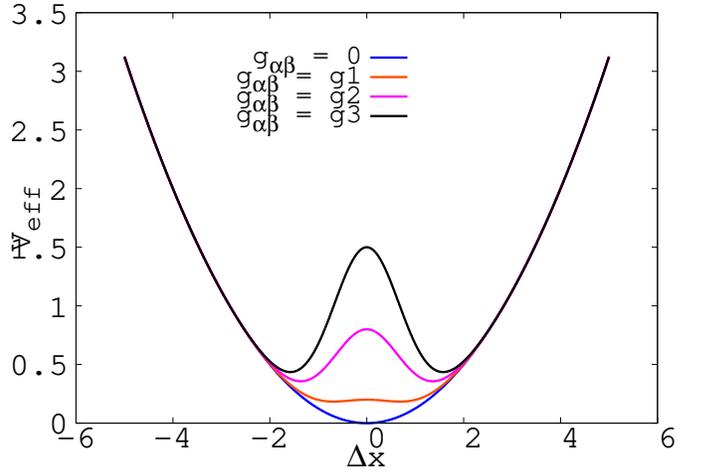}
\caption{Schematic diagram of $V_eff$ with $\Delta x$ for different values of $g_{\alpha\beta}$. When $g_{\alpha\beta}=0$, $V_eff$ remains a simple harmonic potential. As the value of $g_{\alpha\beta}$ increases $g3>g2>g1$, effective potential turns into the double well potential. At sufficient strong value of $g_{\alpha\beta}$ given in condition \ref{eq38} instability occurs at $\delta x=0$ and both of the species occupy the minima of the double well separately.}
\label{fig1}
\end{figure}
 But as we increase $g_{\alpha\beta}$ a barrier grows inside the harmonic potential and for sufficiently large  $g_{\alpha\beta}$ the effective potential becomes a double well potential and in this regime the phase separation occurs with two species occupying the two minima separately.
\section{Discussion:} 

We have generalized the idea of coherent states for single species to the binary species case. The new concept we have applied here can be extended to multi species problem also. In sec II, we have shown the occurrence of instability on coherent states under certain conditions  which is similar to the condition for phase separation already obtained in literature both theoretically and experimentally. Here we have constructed the condition in the context of coherent states. In sec III we have followed the Ehrenfest approach and have shown the occurrence of instability and phase separation under certain condition, effective potential in this regime being in the shape of double well. The transition of the effective potential $(V_{eff})$ from simple harmonic potential to double well due to the inclusion of $g_{\alpha\beta}$ signifies the occurrence of instability and phase separation when there is sufficient interspecies interaction.  

Though these two approaches are somewhat different conceptually, both of them may lead to an instability which can be connected to a phase separation. For the coherent state approach the implications of the instability will be explored in greater detail in future.
\section*{Acknowledgments}
One of the authors, Sukla Pal would like to thank S. N. Bose National Centre for Basic Sciences for the financial support during the work. Sukla Pal acknowledges Harish-Chandra Research Institute also for hospitality and generosity during academic visit.
\appendix
\section{Derivation of Eq.(\ref{eq29}):}
$\frac{d}{dt}\langle x^2\rangle=\frac{d}{dt}\int x^2\psi^{\ast}\psi dx=\int\dot{\psi}_{\alpha}^{\ast}(x)x^2\psi(x)]dx
+\int\dot{\psi}_{\alpha}(x)x^2\psi_{alpha}^{\ast}dx$ and applying Eq. (\ref{eq21}) for both the species we get 
\ber\label{eq27}
\begin{aligned}
\frac{d}{dt}\langle x^2\rangle_{\alpha}=i\int x\Big (\psi_{\alpha}\partial_x\psi_{\alpha}^{\ast}-\psi_{\alpha}^{\ast}\partial_x\psi_{\alpha}\Big)dx\\
\frac{d}{dt}\langle x^2\rangle_{\beta}=i\int x\Big (\psi_{\beta}\partial_x\psi_{\beta}^{\ast}-\psi_{\beta}^{\ast}\partial_x\psi_{\beta}\Big)dx
\end{aligned}
\eer
 We take another time derivative of Eq. (\ref{eq27}) and apply Eq. (\ref{eq21}) again. Then with the application of the property of Gaussian i.e, $\langle x^2\rangle_{\alpha}=x_{0\alpha}^2+W_{\alpha}^2$ for the wave packet of species $\alpha$ and $\langle x^2\rangle_{\beta}=x_{0\beta}^2+W_{\beta}^2$ for that of species $\beta$, Eq. (\ref{eq27}) leads to the followings:
\ber\label{eq28}
\begin{aligned}
x_{0\alpha}\ddot{x_{0\alpha}}+\dot{x_{0\alpha}^2}+W_{\alpha}\ddot{W_{\alpha}}+\dot{W_{\alpha}^2}=\frac{1}{4W_{\alpha}^2}+p_{0\alpha}^2-\Omega^2(W_{\alpha}^2+x_{0\alpha}^2)\\+\frac{g_{\alpha}}{4\sqrt{\pi}}\frac{N_\alpha}{W_{\alpha}}+\frac{g_{\alpha\beta}}{2}\frac{N_{\beta}}{\sqrt{2\pi W^2}}e^{-\frac{\Delta x^2}{2W^2}}\Big[1+2\frac{x_{0\alpha}\Delta x}{W^2}\Big]\\
x_{0\beta}\ddot{x_{0\beta}}+\dot{x_{0\beta}^2}+W_{\beta}\ddot{W_{\beta}}+\dot{W_{\beta}^2}=\frac{1}{4W_{\beta}^2}+p_{0\beta}^2-\Omega^2(W_{\beta}^2+x_{0\beta}^2)\\+\frac{g_{\beta}}{4\sqrt{\pi}}\frac{N_\beta}{W_{\beta}}+\frac{g_{\alpha\beta}}{2}\frac{N_{\alpha}}{\sqrt{2\pi W^2}}e^{-\frac{\Delta x^2}{2W^2}}\Big[1+2\frac{x_{0\beta}\Delta x}{W^2}\Big]
\end{aligned}
\eer
With the help of Eq. (\ref{eq25}) and Eq. (\ref{eq26}) we reach into Eq. (\ref{eq29}).

\end{document}